\documentclass[aps,nofootinbib,amsmath,prc,showpacs,groupedaddress,12pt,superscriptaddress,notitlepage]{revtex4-1}
\usepackage[paperwidth=210mm,paperheight=297mm,centering,hmargin=2.1cm,tmargin=2.3cm,bmargin=2.8cm]{geometry}

\usepackage{amsfonts}
\usepackage{amsmath}
\usepackage{amssymb}
\usepackage{epsfig}
\usepackage{latexsym}
\usepackage{paralist}
\usepackage{fancyhdr}
\usepackage{graphicx}
\usepackage[vcentermath]{youngtab}
\usepackage{young}
\usepackage{etex}
\usepackage{braket}
\usepackage{float}
\usepackage{slashed}
\usepackage{xcolor}
\usepackage{soul}
\usepackage{dcolumn}
\usepackage{physics}
\usepackage{amsmath}

\def\bea{\begin{eqnarray}}
\def\eea{\end{eqnarray}}
\def\be{\begin{equation}}
\def\ee{\end{equation}}
\def\ba{\begin{array}}
\def\ea{\end{array}}

\def\Tr{\mbox{Tr}}

\def\ep{\epsilon}

\def\beq{\begin{eqnarray}}    
\def\eeq{\end{eqnarray}}       

\def\D{\Delta_g}       

\def\al{\alpha}
\def\be{\beta}

\def\ga{\gamma}

\def\ep{\epsilon}

\def\Ga{\Gamma}

\def\xibar{\left(\xi-\frac{1}{6} \right)}

\def\cB{{\cal B}}
\def\cC{{\cal C}}

\usepackage{mathpazo}

\begin{document}

\title{Vacuum effective actions and mass-dependent \\ renormalization in curved
space\footnote{%
Prepared for the special issue of Universe collecting the contributions
to the workshop ``Quantum Fields—from Fundamental Concepts to Phenomenological Questions'', Mainz 26-28 September 2018}
}

\author{Sebasti{\'a}n~A.~Franchino-Vi{\~n}as}
\email{sa.franchino@uni-jena.de}
\affiliation{Theoretisch-Physikalisches Institut,
Friedrich-Schiller-Universit\"{a}t Jena, Max-Wien-Platz 1,
07743 Jena, Germany}

\author{Tib{\'e}rio~de~Paula~Netto}
\email{tiberiop@fisica.ufjf.br}
\affiliation{Departamento de F{\'i}sica, ICE, Universidade Federal de
Juiz de Fora, Juiz de Fora, 36036-100, Minas Gerais, Brazil}

\author{Omar~Zanusso}
\email{omar.zanusso@uni-jena.de}
\affiliation{Theoretisch-Physikalisches Institut,
Friedrich-Schiller-Universit\"{a}t Jena, Max-Wien-Platz 1,
07743 Jena, Germany}

\begin{abstract}
\noindent
We review past and present results on the non-local form-factors of the effective action of semiclassical gravity
in two and four dimensions computed by means of a covariant expansion of the heat kernel up to the second order in the curvatures.
We discuss the importance of these form-factors in the construction of mass-dependent beta functions for the Newton's constant
and the other gravitational couplings.
\end{abstract}

\maketitle


\section{Introduction}\label{sect:introduction}

The Appelquist-Carazzone theorem implies that quantum effects induced by the integration
of a massive particle are suppressed when studied at energies smaller than a threshold
set by the particle's mass \cite{AC}. The suppression mechanism has been well understood both quantitatively
and qualitatively in flat space. From a renormalization group (RG) perspective it is convenient to
adopt a mass-dependent renormalization scheme, which shows that the running of couplings that are induced by
the integration of massive fields is suppressed below the mass threshold.
Extensions of the above statements to curved space have been developed only more recently
because of the additional difficulties in preserving covariance.
In curved space it is convenient to compute the vacuum effective action, also known as the semiclassical action,
which is the effective metric action induced by the integration of matter fields.
If the effective action is computed correctly, the decoupling mechanism can be studied covariantly through
the use of opportune form-factors among the curvatures. These form-factors are in fact covariant functions of
the Laplacian, both in two- \cite{Ribeiro:2018pyo} and four- \cite{apco,fervi,BuGui,Franchino-Vinas:2018gzr} dimensional curved space.

The simplest way to compute the necessary form-factors and maintain covariance is through the use of the heat kernel expansion \cite{bavi85}.
For our purposes it is convenient to adopt a curvature expansion, which resums the covariant derivatives acting on the curvatures as the non-local form-factors \cite{bavi87,bavi90}.
More precisely, it proves essential to use a heat kernel expansion which resums the total derivative terms constructed
by an arbitrary power of the Laplacian acting on a single curvature scalar $R$ \cite{Codello:2012kq}.
This paper reviews the recent developments on the use of these boundary terms to investigate the decoupling
of the Newton's constant \cite{Ribeiro:2018pyo,Franchino-Vinas:2018gzr}.
We believe that these develpments might be useful in the broader context of developing
non-local effective actions which have useful phenomenological implications.
Among these we include the anomaly induced inflation models \cite{susykey,Shocom,asta},
even though they are not sufficient for deriving Starobinsky's inflation purely from quantum corrections \cite{star,star83}.
Our results might pave the way to the construction of a field theoretical model \cite{StabInstab}.
More generally, renormalization-group-running Newton's and cosmological constants
could have measurable implications in both cosmology \cite{CC-Gruni} and astrophysics
\cite{RotCurves}. For this purpose, runnings developed using spacetimes of non-zero
constant curvature are a first step \cite{DCCrun,Verd},
which have to be reconciled with the same runnings
that are obtained in the modified minimal subtraction ($\overline{\rm MS}$) scheme \cite{nelpan82,buch84,book}.

Focussing our attention on phenomenologically interesting effective actions
it is important to mention that non-local actions are promising candidates to
describe dark energy \cite{Maggiore,CC-Gruni,DCCrun,Codello:2015pga},
as well as satisfying templates
to reconstruct the effective action induced by dynamical triangulations or asymptotic safety \cite{Knorr:2018kog}.
The applications might even extend to Galileon models, especially if promoted to their covariant counterparts \cite{Codello:2012dx,Brouzakis:2013lla}
with form-factors that act also on extrinsic curvatures \cite{Codello:2011yf}.
The most recent results on the renormalization of Newton's constant in a massive scheme point to
the necessity of connecting the renormalization of the operators $R$, $\Box R$ and $R^2$ \cite{Ribeiro:2018pyo,Franchino-Vinas:2018gzr},
and that the couplings could be generalized to $\Box$-dependent functions,
a fact which is reminiscent of previous analyses by Avramidi \cite{Avramidi:2007zz} and by Hamber and Toriumi \cite{Hamber:2010an,Hamber:2011kc}.
In this respect, the relations among the non-local form-factor of the above terms in the semiclassical
theory has already been emphasized in \cite{anom2003}.

This paper reviews the recent results on the mass-dependent renormalization of the Newton's constant
induced by the integration of massive matter fields
in two \cite{Ribeiro:2018pyo} and four \cite{Franchino-Vinas:2018gzr} dimensions,
complementing the latter with results that previously appeared in \cite{apco,fervi,BuGui}.
The outline of this review is as follows: In section \ref{sect:mass-dependent-schemes} we briefly describe the decoupling
of the electron's loops in electrodynamics and connect it with the computation of the QED semiclassical action.
In section \ref{sect:effective-action} we introduce the covariant representation of the effective action
that underlies this work.
In sections \ref{sect:nonlocal-two} and \ref{sect:nonlocal-four} we apply our formalism
to two- and four-dimensional curved space respectively. We concentrate on scalar, Dirac and Proca fields in both cases.
In section \ref{sect:uv-structure} we briefly describe the general structure of the effective action
and make some general statement on its ultraviolet structure.
In section \ref{sect:scheme} we speculate that our formalism could have untapped potential
for expressing results of the asymptotic safety conjecture \cite{Reuter:1996cp,books} by making
the case of scheme independence.
The appendices \ref{sect:heat-kernel} and \ref{sect:further} contain mathematical details on the heat kernel
and on the geometrical curvatures that would have otherwise burdened the main text.

\section{Mass-dependent schemes}\label{sect:mass-dependent-schemes}

In this section we outline our strategy to find explicit predictions of the Appelquist-Carazzone theorem in the simpler setting of quantum electrodynamic (QED) in flat space.
In particular, we take this opportunity to bridge the gap between the more traditional approach and a fully covariant method.
We begin by considering the regulated one-loop vacuum polarization tensor of QED in $d=4-\epsilon$ dimensions
\begin{equation}\label{eq:vacuum-polarization}
 \begin{split}
 \frac{e^2}{2\pi^2}\left(q^2 g_{\mu\nu}-q_\mu q_\nu\right)\left[ -\frac{1}{3\bar{\epsilon}}  + \int_0^1{\rm d}\alpha \, \alpha(1-\alpha) \ln\left(\frac{m^2+\alpha(1-\alpha)q^2}{m^2}\right)\right]\,,
 \end{split}
\end{equation}
in which $q_\mu$ is the momentum of the external photon lines and $m^2$ is the square mass of the electron that is integrated in the loop.
In the modified minimal subtraction scheme ($\overline{\rm MS}$) one subtracts the contribution proportional to $\frac{1}{\bar{\epsilon}}$ which includes the dimensional pole as well as some finite terms
\begin{equation}
 \begin{split}
 \frac{1}{\bar{\epsilon}} 
 &
 = \frac{1}{\ep} + \frac{1}{2}\ln \left(\frac{4\pi\mu^2}{m^2}\right) 
- \frac{\ga}{2}
\end{split}
\end{equation}
($\ga\simeq 0.5$ is the Euler's constant), so that the resulting finite polarization is
\begin{equation}
 \begin{split}
 \frac{e^2}{2\pi^2}\left(q^2 g_{\mu\nu}-q_\mu q_\nu \right)\int_0^1{\rm d}\alpha \, \alpha(1-\alpha) \ln\left(\frac{m^2+\alpha(1-\alpha)q^2}{m^2}\right)\,.
 \end{split}
\end{equation}
Customarily, the regularization procedure introduces a scale $\mu$ and the dependence of the renormalized constant $e(\mu)$ on this scale is encoded
in the beta function
\begin{equation}
 \begin{split}
 \beta^{\overline{\rm MS}}_{e}
 = \frac{e^3}{12\pi^2}\,,
 \end{split}
\end{equation}
which comes essentially from the coefficient of the subtracted pole times $\frac{e}{2}$ \cite{BuGui}.
Notice that we labelled the beta function with $\overline{\rm MS}$ so that it is clear that we used the modified minimal subtraction scheme to compute it.

An alternative to the $\overline{\rm MS}$ scheme would use some other scale to subtract the divergence,
this new choice generally results in a mass-dependent scheme if the new scale is not $\mu$.
If we choose as new scale $q=\left|q_\mu\right|$, a different beta function can be computed by acting
on the right term between the brackets in \eqref{eq:vacuum-polarization} with $\frac{e}{2}p\partial_p$ \cite{apco}
resulting in
\begin{equation}\label{eq:beta-qed}
 \begin{split}
 \beta_{e}
 = \frac{e^3}{2\pi^2}\int_0^1{\rm d}\alpha \, \alpha(1-\alpha) \frac{\alpha(1-\alpha)q^2}{m^2+\alpha(1-\alpha)q^2}\,.
 \end{split}
\end{equation}
The new beta function explicitly depends on the mass of the electron, besides the scale $q$, thus allowing us to distinguish the following two limits
\begin{equation}
 \begin{split}
  \beta_{e} & \simeq 
           \begin{cases}
             \frac{e^3}{12\pi^2} & \qquad {\rm for} \quad q^2 \gg m^2 \,;\\
             \frac{e^3}{60\pi^2} \frac{q^2}{m^2} & \qquad {\rm for} \quad q^2 \ll m^2\,.
           \end{cases}
 \end{split}
\end{equation}
The physical interpretation of the above results goes as follows: in the ultraviolet, which corresponds to energies $q^2$ much bigger than the electron's mass,
the beta function coincides with its $\overline{\rm MS}$ counterpart which is a universal result
at high energies.\footnote{%
This happens because the scale $\mu$
of dimensional regularization, which we use to subtract the poles, can be interpreted as a very high energy scale which is bigger than any other scale in the theory
and in particular bigger than the electron's mass.}
Instead in the infrared, which corresponds to energies $q^2$ smaller than the electron's mass,
the electron in the loop hits the mass threshold and effectively stops propagating. This results in a contribution to the renormalization group (RG) that goes
to zero quadratically with the energy $q$.
This latter effect is predicted in general terms by the Appelquist-Carazzone theorem
and can be observed in any quantum field theoretical computation that involves massive particles
propagating in the loops.

As anticipated, in this contribution we generalize similar results to several types of massive fields in two- and four-dimensional curved spacetimes.
In dealing with curved space it is convenient to have results that are always manifestly covariant \cite{Donoghue:2017pgk}.
In order to achieve manifest covariance we are going to present an effective-action-based computation which can be done using the heat kernel methods
described in appendix \ref{sect:heat-kernel}, and illustrate how the above results are derived from a covariant effective action.
Using non-local heat kernel methods one finds that
the renormalized contributions to the vacuum effective action of QED become
\begin{equation}
 \begin{split}
  \Gamma[A] = 
  \frac{1}{4}\int {\rm d}^4y\, F_{\mu\nu}F^{\mu\nu}
   - \frac{e^2}{8\pi^2} \int {\rm d}^4x \, F_{\mu\nu}\left\{
  \int_0^1 {\rm d}\alpha \, \alpha(1-\alpha)\ln \left(\frac{m^2+\alpha(1-\alpha) \Delta}{4\pi\mu^2}\right)
  \right\} F^{\mu\nu}\,,
 \end{split}
\end{equation}
in which $\Delta=-\partial_x^2$ is the Laplacian operator in flat space and $F_{\mu\nu}=\partial_\mu A_\nu-\partial_\nu A_\mu$ is the Abelian curvature tensor \cite{Codello:2015oqa}.
It should be clear that the non-local form-factor appearing
between the two copies of $F_{\mu\nu}$ is a covariant way of writing \eqref{eq:vacuum-polarization}
in which the momentum scale $q^2$ comes from Fourier transformation of the differential operator $\Delta$.

Using this latter observation, one could proceed to the computation
of the mass-dependent beta function by ``undoing'' the covariantization and by extracting the form-factor to obtain \eqref{eq:vacuum-polarization}.
In practical computations we replace $\Delta$ with the square
of the new reference scale $q^2$ and apply the derivatives with respect to $q$ as outlined before \cite{Goncalves:2009sk},
thus following closely the steps that lead to \eqref{eq:beta-qed}. This latter strategy of identifying the relevant scale
with the covariant Laplacians of the effective action's form-factors can be easily applied to curved space,
in which there are more curvature tensors besides $F_{\mu\nu}$ and therefore more couplings,
and it will prove fundamental for the rest of this review.

\section{Heat kernel representation of the effective action in curved space}\label{sect:effective-action}

We now concentrate our attention to a $D$-dimensional spacetime in which the dimensionality can be either $D=2$ or $D=4$.
We assume that the spacetime is equipped with a classical torsionless Euclidean metric $g_{\mu\nu}$, which for practical purposes can be assumed
to come from the Wick rotation of a Lorentzian metric.
Our task is to compute the vacuum effective actions for the classical metric induced by the integration of massive matter fields.
If we limit our interest to fields of spin up to one, we must consider scalars, spinors and vectors,
which is why we consider the following bare actions
\begin{equation}\label{eq:bare-actions}
 \begin{split}
 S_{\rm s}[\varphi] &= \frac{1}{2}\int {\rm d}^Dx\sqrt{g} \left( g^{\mu\nu}\partial_\mu\varphi\partial_\nu\varphi + m_{\rm s}^2\varphi^2+ \xi \varphi^2 R \right) \\
 S_{\rm f}[\psi] &= \int {\rm d}^Dx \sqrt{g}\, \overline{\psi}\left(\slashed D+m_{\rm f} \right)\psi \\
 S_{\rm p}[A] &= \int {\rm d}^Dx\sqrt{g} \left(-\frac{1}{4} F_{\mu\nu} F^{\mu\nu} + \frac{m_{\rm v}^2}{2} A_\mu A^\mu\right) 
 \end{split}
\end{equation}
in which we defined $\slashed D = \gamma^a e^\mu{}_a D_\mu$, $D_\mu = \partial_\mu + S_\mu$ with $S_\mu$ the spin-$\frac{1}{2}$ connection,
$F_{\mu\nu} = \nabla_\mu A_\nu -\nabla_\nu A_\mu$ and $R$ is the scalar curvature. The action $S_{\rm s}[\varphi]$ represents a non-minimally coupled
free massive scalar field, while $S_{\rm f}[\psi]$ and $S_{\rm p}[A]$ represent minimally coupled massive Dirac spinors and massive Proca vectors respectively.

Given that the matter fields are quadratic, the one-loop effective action corresponds to the full integration of the path-integral
and captures a physical situation in which the matter interactions are weak.
If we have $n_{\rm s}$ scalars, $n_{\rm f}$ Dirac spinors and $n_{\rm p}$ Proca vectors of equal masses per spin,
the full effective action is additive in its sub-parts
\beq
\Ga[g] &=& n_{\rm s} \Gamma_{\rm s}[g] + n_{\rm f} \Gamma_{\rm f}[g] + n_{\rm p} \Gamma_{\rm p}[g]\,,
\eeq
in which the single contributions can be easily obtained from a standard path-integral analysis
\begin{equation}\label{eq:functional-traces}
 \begin{split}
 &
 \Gamma_{\rm s}[g] 
 = \frac{1}{2} \Tr_{\rm s} \ln \left( \D + \xi R +m_{\rm s}^2\right) , 
\\
 & 
 \Gamma_{\rm f}[g] = -\Tr_{\rm f} \ln \left(\slashed D+m_{\rm f}\right) , 
 \\
 &
 \Gamma_{\rm p}[g] = \frac{1}{2}\Tr_{\rm v} \ln\left( \delta_\mu^\nu \D 
 +\nabla_\mu\nabla^\nu + R_\mu{}^\nu + \delta_\mu^\nu m_{\rm v}^2\right) ,
 \end{split}
\end{equation}
and we defined the curved space Laplace operator $\D= - \nabla^2 = -\nabla_\mu\nabla^\mu= -g_{\mu\nu}\nabla^\mu\nabla^\nu$.

One notices that $\Gamma_{\rm s}[g]$ is a functional trace of an operator of Laplace-type, and therefore can be dealt with
using standard heat kernel methods. The same is not true for the other two traces, but it is a well-known fact that
we can manipulate them to recover a Laplace-type operator.
For the Dirac fields it is sufficient to recall that the square $\left({\rm i}\slashed D\right)^2=\D + \frac{R}{4}$, which implies
\begin{equation}
 \begin{split}
 \Gamma_{\rm f}[g] &= -\frac{1}{2} \Tr_{\rm f} \ln\left[\left(\slashed D+m_{\rm f}\right)^2\right]
 =-\frac{1}{2} \Tr_{\rm f} \ln \left( \D + \frac{R}{4} + m_{\rm f}^2\right) \,,
 \end{split}
\end{equation}
if we assume a positive bounded spectrum for the Dirac operator.
A more involved manipulation can be done to the Proca's functional trace \cite{bavi85,Ruf:2018vzq} and it results in
\begin{equation}
 \begin{split}
 \Gamma_{\rm p}[g] &= \frac{1}{2} \Tr_{\rm v} \ln\left( \D+{\rm Ric}+m_{\rm v}^2\right) - \frac{1}{2} \Tr_{\rm s} \ln\left(\D +m_{\rm v}^2\right)\,.
 \end{split}
\end{equation}
The physical interpretation of the above difference is that a Proca field can be understood
as a vector degree of freedom which is integrated in the first trace, minus one single scalar ghost which is integrated in the second trace,
for a total of one degree of freedom in $D=2$ and three degrees of freedom in $D=4$.
Recall now that the functional trace of a Maxwell's $U(1)$ gauge field, which naively could be understood as massless Proca vector,
includes the subtraction of \emph{two} ghost degrees of freedom, which is one more than the Proca's.
This shows that the naive limit $m_{\rm v}\to 0$ does not actually recover a Maxwell field, but rather it is discontinuous.

A simple glance at all the above traces shows that, modulo overall constants, we are generally interested in functional traces of Laplace-type operators in the form
\begin{equation}\label{eq:effective-action-trlog-generic}
 \begin{split}
 \Ga[g] &= \frac{1}{2} \Tr \ln \left(\D +E +m^2\right)\,,
 \end{split}
\end{equation}
in which we trace over the opportune degrees of freedom. The general endomorphism $E=E(x)$ acts on the field's bundle
and it is assumed to be arbitrary, so that by taking the opportune form we obtain the result of either of the above traces.
Let us collectively denote the general Laplace-type operator ${\cal O}= \D+ E$ and its heat kernel ${\cal H}_D(s;x,x')$,
in which we keep the subscript $D$ as a reminder of the spacetime dimension for later use.
Following appendix \ref{sect:heat-kernel} we use the heat kernel to represent \eqref{eq:effective-action-trlog-generic} as
\begin{equation}
 \begin{split} \label{eq:effective-action-divergent}
 \Ga[g] &= -\frac{1}{2} {\rm tr} \int_0^\infty \frac{{\rm d}s}{s} \int{\rm d}^Dx ~ {\rm e}^{-sm^2} {\cal H}_D(s;x,x)\,,
 \end{split}
\end{equation}
%
in which the bi-tensor ${\cal H}_D(s;x,y)$ is the solution of the heat kernel evolution equation in $D$-dimensions.
The effective action \eqref{eq:effective-action-divergent} is generally an ultraviolet divergent functional: divergences appear as poles in the integration
of the $s$ variables at $s=0$ because $s$ is conjugate to the square of a momentum.
The leading power of the heat kernel is $s^{-D/2}$ and, after expanding in powers of $s$,
one expects a finite number poles for the first few terms of this expansion.
In particular, the first two terms will contain divergences for $D=2$, or the first three for $D=4$ (see also below).

We regularize divergences by analytic continuation of the dimensionality to $d=D-\epsilon$. Since in curved space the dimensionality can appear
in a multitude of ways (such as $g_\mu{}^\mu$) we have to be careful in our choice for the analytic continuation.
We choose to continue only the leading power of the heat kernel, thus promoting ${\cal H}_D(s;x,x)\to{\cal H}_d(s;x,x)$,
while at the same time keeping all geometrical objects in $D$ dimensions (implying, for example, that $g_\mu{}^\mu=D$ and \emph{not} $g_\mu{}^\mu=d$).
This choice is probably the simplest that one can make, but we should stress that any other choice differs from this one
by finite terms which do not change the predictions of the renormalized effective action.
After our continuation to $d$ dimensions the trace becomes
\begin{equation}
 \begin{split} \label{eq:effective-action-regularized}
 \Ga[g] &= -\frac{\mu^{\epsilon}}{2} \tr \int_0^\infty \frac{{\rm d}s}{s} \int{\rm d}^Dx ~ {\rm e}^{-sm^2} {\cal H}_d(s;x,x)\,,
 \end{split}
\end{equation}
in which we have also introduced a reference scale $\mu$ to preserve the mass dimension of all quantities
when leaving $D$ dimensions, and the label $d$ of the heat kernel is a reminder of the continuation $s^{-D/2}\to s^{-d/2}$ \cite{bro-cass}.

Before concluding this section we find convenient to introduce some further definition. When studying the renormalization group
it is sometimes useful to consider dimensionless variables. At our disposal we have the renormalization group scale $q$
which is related to $\Delta_g \leftrightarrow q^2$ as discussed in section \ref{sect:mass-dependent-schemes},
and a mass $m$ which collectively denotes the species' masses introduced before.
For us it is natural to give every dimensionful quantity in units of the mass $m$,
which leads to the following dimensionless operators
\begin{equation}\label{eq:dimensionless-operators}
 z = \frac{\D}{m^2}\,,
 \qquad
 a = \sqrt{\frac{4z}{4+z}}\,,
 \qquad
Y = 1-\frac{1}{a} \ln\left|{\frac{1+a/2}{1-a/2}}\right|\,.
\end{equation}
We will also denote by $\hat{q}^2=q^2/m^2$ the dimensionless RG scale (the RG scale in units of the mass),
which is related to $z \leftrightarrow \hat{q}^2$ according to the discussion of section \ref{sect:mass-dependent-schemes}.
We will not adopt further symbols for the operators $a$ and $Y$ after the identification,
which means that from the point of view of the RG they will be functions of the ratio $\hat{q}^2=\frac{q^2}{m^2}$
and therefore change as a function of the energy.

\section{Renormalized action in two dimensions}\label{sect:nonlocal-two}

In $D=2$ the only independent curvature tensor is the Ricci scalar $R$ if there are no further gauge connections.
We therefore choose to parametrize the most general form that a regularized effective action can take as
\begin{equation} \label{eq:effective-action-2d}
 \begin{split}
\Ga[g] &=
\Ga_{\rm loc}[g]
 + \frac{1}{4\pi}\int {\rm d}^2 x \sqrt{g}\, \cB(z) R
 - \frac{1}{96\pi}\int {\rm d}^2 x \sqrt{g}\, 
 R\, \frac{\cC(z)}{\D} \, R\,.
 \end{split}
\end{equation}
The part $\Ga_{\rm loc}[g]$ is a local function of the curvatures and as such contains the divergent
contributions which require the renormalization of both zero point energy and coefficient of the scalar curvature.
These two divergences correspond to the leading $s^{-d/2}$ and subleading $s^{-d/2+1}$ (logarithmic)
powers of the expansion of the heat kernel. Starting from the terms that are quadratic in the scalar curvature
the parametric $s$ integration becomes finite.

The dimensional divergences that appear in $\Ga_{\rm loc}[g]$ can be renormalized by opportunely choosing two counterterms
up to the first order in the curvatures. Consequently, after the subtraction of the divergences, the local part of the renormalized action contains
\begin{equation}
 \begin{split}
 S_{\rm ren}[g]
 &= \int {\rm d}^2 x \sqrt{g}\,\left\{b_0 + b_1 R\right\}
 \end{split}
\end{equation}
in which the couplings $b_0$ and $b_1$ are related to the two-dimensional cosmological and Newton's constants.
A popular parametrization of the Einstein-Hilbert action in two dimensions is $b_0=\Lambda$ and $b_1=-G^{-1}$,
in which $\Lambda$ and $G$ are the two-dimensional cosmological and Newton's constants respectively.
The $\overline{\rm MS}$ procedure generates perturbative beta functions for the renormalized couplings
which we denote with $\beta^{\overline{\rm MS}}_{b_0}$ and $\beta^{\overline{\rm MS}}_{b_1}$ and which depend on the specific matter content.

The non-local part of \eqref{eq:effective-action-2d} is also very interesting for our discussion.
If the critical theory is conformally invariant, then we know that it contains the pseudo-local Polyakov action
\begin{equation}\label{eq:polyakov-action}
 \begin{split}
 S_{\rm P}[g]
 &= - \frac{c}{96\pi}\int {\rm d}^2 x \sqrt{g}\, 
 R\, \frac{1}{\D} \, R\,,
 \end{split}
\end{equation}
in which we introduced the central charge of the conformal theory $c$ \cite{Barvinsky:2004he}. The Polyakov action accounts for the violations of the conformal symmetry from the measure of the path integral
at the quantum level \cite{Codello:2014wfa}.
The central charge counts the number of degrees of freedom of the model and
it is generally understood as a property of the fixed points of the renormalization group, which in general means that $c=c(g^*)={\rm const.}$
for $g^*$ some fixed point coupling(s).

Since the Polyakov action is not required for the subtraction of any divergence we could deduce that the ${\overline{\rm MS}}$ scheme does not generate a flow
for the central charge, or alternatively $\beta^{\overline{\rm MS}}_{c}=0$. This latter property is in apparent contradiction to Zamolodchikov's theorem that states that $\Delta c\leq 0$
along the flow, but the contradiction is qualitatively resolved by understanding that the ${\overline{\rm MS}}$ scheme captures only the far ultraviolet of the RG flow.
A comparison of \eqref{eq:polyakov-action} with \eqref{eq:effective-action-2d} suggests the interpretation
of the function $\cC(z)$ as a RG-running central charge in our massive scheme, recalling that $z$ is the square of our RG scale in units of the mass.

Our framework makes a quantitative connection with Zamolodchikov's theorem:
the non-local part of the effective action is parametrized by the functions $\cB(z)$ and $\cC(z)$, which are both dimensionless functions
of the dimensionless argument $z$. Simple intuition allows us to interpret $\cB(z)$ as a non-local generalization of the Newton's constant,
while we suggest to interpret $\cC(z)$ as a generalization of the central charge under the correct conditions (see below).
In all applications below we observe that $\Delta \cC \leq 0$ for flows connecting known conformal theories, in agreement with the theorem \cite{Zamolodchikov:1986gt}.

As discussed in section \ref{sect:mass-dependent-schemes}, we introduce the momentum scale $q$ and its dimensionless counterpart $\hat{q}=q/m$. Setting the momentum scale from $z=\hat{q}^2$
and interpreting the coefficient of $R$ as a scale dependent coupling
we define the non-local beta function of $b_1$
\begin{equation}
 \begin{split}
\beta_{b_1} &= q\frac{\partial}{\partial q} \frac{\cB(z)}{4\pi}= \hat{q}\frac{\partial}{\partial \hat{q}} \frac{\cB(z)}{4\pi} = \frac{z}{2\pi} \cB'(z)\,,
 \end{split}
\end{equation}
in which we used a prime to indicate a derivative with respect to the argument.
Analogously we push the interpretation of the derivative of $\cC(z)$ as a running central charge
\begin{equation}\label{eq:running-central-charge}
 \begin{split}
 \beta_c &= q\frac{\partial}{\partial q} \cC(z) = 2  z \,\cC'(z)\,.
 \end{split}
\end{equation}
Again we stress that this latter flow is expected to be negative for trajectories connecting two conformal field theories to comply with Zamolodchikov's theorem.

In agreement with general arguments, we see that the UV limit of the non-local beta functions reproduce the standard ${\overline{\rm MS}}$ results.
Specifically we have that the running of $b_1$ reproduces the $\overline{\rm MS}$ result at high energies
\begin{equation}
 \begin{split}
 \beta_{b_1} &= \beta^{\overline{\rm MS}}_{b_1} + {\cal O}\left(\frac{m^2}{q^2}\right) \qquad {\rm for}\quad q^2\gg m^2\,.
 \end{split}
\end{equation}
We also see that the non-local generalization of the central charge is related to the central charge itself in the same limit
\begin{equation}
 \begin{split}
  \cC(z) = c + {\cal O}\left(\frac{m^2}{q^2}\right) \qquad {\rm for}\quad q^2\gg m^2\,.
 \end{split}
\end{equation}
This latter property seems to be always true if $c$ is interpreted as the number of degrees of freedom of the theory.
In particular it is true for the case of the Proca field which is not conformally invariant like the massless minimally coupled scalar or the massless Dirac field.
We will see in the next sections that $c=1$ for scalars with $\xi=0$, $c=1/2$ for spinors, and $c=1$ for Proca fields in two dimensions.
All the explicit expressions for the functions $\cB(z)$, $\cC(z)$ and their derivatives are given in the next three subsections.

\subsection{Non-minimally coupled scalar field in two dimensions}\label{sect:scalar2d}

We now give all the terms needed for the scalar field trace appearing in \eqref{eq:functional-traces} in $D=2$.
As a template to assemble all terms we refer to \eqref{eq:effective-action-2d}.
The local part of the effective action is
\begin{equation}\label{eq:effective-action-2d-scalar-local}
 \begin{split}
 \Gamma_{\rm loc}[g] &=
 \frac{1}{4\pi}\int {\rm d}^2 x \sqrt{g}\, \Bigl\{
 \left(\frac{1}{\bar{\epsilon}}+\frac{1}{2}\right)m^2
  +\left(\xi-\frac{1}{6}\right)\frac{1}{\bar{\epsilon}} R
 \Bigr\}\,,
 \end{split}
\end{equation}
which has poles in both terms as expected.
The non-local part of \eqref{eq:effective-action-2d} is captured by the functions
\begin{equation}
 \begin{split}
 \cB(z)
 &= \frac{1}{36}+\left(\xi-\frac{1}{4}\right)Y+\frac{Y}{3a^2}
 \\
 \cC(z)
 &=
 -\frac{1}{2}-\frac{6Y}{a^2}-12\left(\xi-\frac{1}{4}\right)Y+6\left(\xi-\frac{1}{4}\right)^2(1-Y)\,,
 \end{split}
\end{equation}
in which we use the notation \eqref{eq:dimensionless-operators}. From the non-local functions we can derive
the mass-dependent beta function
\begin{equation}
 \begin{split}
 \beta_{b_1} &= \frac{z}{2\pi} \cB'(z)
 = \frac{1}{2\pi}\Bigl\{
 -\frac{1}{24}-\frac{Y}{2a^2}-\frac{1}{2}\left(\xi-\frac{1}{2}\right)Y-\frac{1}{8}\left(\xi-\frac{1}{4}\right)(1-Y)a^2
 \Bigr\}\,.
 \end{split}
\end{equation}
The beta function in the mass-dependent scheme displays two limits
\begin{equation}
 \begin{split}
 \beta_{b_1} &= \begin{cases}
            \frac{1}{4\pi}\left(\frac{1}{6}-\xi\right) +{\cal O}\left(\frac{m^2}{q^2}\right) & \qquad {\rm for} \quad q^2 \gg m^2 \,; \\
            \frac{1}{24\pi}\left(\frac{1}{5}-\xi \right) \frac{q^2}{m^2} +{\cal O}\left(\frac{q^2}{m^2}\right)^2 & \qquad {\rm for} \quad q^2 \ll m^2\,.
           \end{cases}
 \end{split}
\end{equation}
The low energy limit shows a realization of the Appelquist-Carazzone theorem for which the Newton's constant stops running below
the threshold determined by the mass with a quadratic damping factor.
The high energy limit shows instead that $\beta_{b_1}$ reduces to minus the coefficient of $R$'s divergent term in \eqref{eq:effective-action-2d-scalar-local}
and thus to its $\overline{\rm MS}$ counterpart.
One can explicitly check that $\beta_{c}$ defined as in \eqref{eq:running-central-charge} is positive as a function of $z$ if $\xi=0$, meaning that $\Delta \cC\leq 0$
from the UV to the IR.
For practical purposes we are interested in
\begin{equation}
 \begin{split}
 \cC(z)&= 
           \begin{cases}
            1  -12 \xi +12 \xi^2 \ln \left(\frac{q^2}{m^2}\right) +{\cal O}\left(\frac{m^2}{q^2}\right) & \qquad {\rm for} \quad q^2 \gg m^2 \,; \\
            0 +{\cal O}\left(\frac{q^2}{m^2}\right) & \qquad {\rm for} \quad q^2 \ll m^2\,.
           \end{cases}
 \end{split}
\end{equation}
Notice in particular that $\cC(\infty)=1$ for $\xi=0$, which is the central charge of a single minimally coupled free scalar
and is expected from the general result $\Delta \cC = c_{\rm UV} - c_{\rm IR}=1$ under the normalization $c_{\rm IR}=1$.
The interpretation of this result is that for $\xi=0$ the RG trajectory connects a theory with $c=1$ with the massive theory with $c=0$
that lives in the infrared.

\subsection{Dirac field in two dimensions}\label{sect:dirac2d}

Here we report all the terms needed for the Dirac field trace appearing in \eqref{eq:functional-traces} in $D=2$.
The template is again \eqref{eq:effective-action-2d} and
we denote by $d_\gamma$ the dimensionality of the Clifford algebra,
which factors in front of all formulas (see also the discussion at the end of appendix \ref{sect:further}).
The local part of the effective action is
\begin{equation}
 \begin{split}
 \Gamma_{\rm loc}[g] &=
 \frac{d_\gamma}{4\pi}\int {\rm d}^2 x \sqrt{g}\, \Bigl\{
 -\left(\frac{1}{\bar{\epsilon}}+\frac{1}{2}\right)m^2
  -\frac{1}{12}\frac{1}{\bar{\epsilon}} R
 \Bigr\}\,,
 \end{split}
\end{equation}
which has poles in both terms as expected.
The non-local part of \eqref{eq:effective-action-2d} is captured by the functions
\begin{equation}
 \begin{split}
 \cB(z)
 = d_\gamma\left\{-\frac{1}{36}-\frac{Y}{3a^2}\right\}\,,
 &\qquad
 \cC(z)
 =
 d_\gamma\left\{\frac{1}{2}-\frac{3}{2} Y + \frac{6Y}{a^2}\right\}\,.
 \end{split}
\end{equation}
From the first non-local function we can derive the mass-dependent beta function
\begin{equation}
 \begin{split}
 \beta_{b_1} = 
 &= \frac{d_\gamma}{2\pi}\left\{\frac{1}{24}-\frac{Y}{8} + \frac{Y}{2a^2}\right\}
 \end{split}
\end{equation}
which displays two limits
\begin{equation}
 \begin{split}
 \beta_{b_1} &= \begin{cases}
            \frac{d_\gamma}{24\pi}\frac{1}{2} +{\cal O}\left(\frac{m^2}{q^2}\right) & \qquad {\rm for} \quad q^2 \gg m^2 \,; \\
           \frac{d_\gamma}{24\pi} \frac{1}{20} \frac{q^2}{m^2} +{\cal O}\left(\frac{q^2}{m^2}\right)^{\frac{3}{2}} & \qquad {\rm for} \quad q^2 \ll m^2\,.
           \end{cases}
 \end{split}
\end{equation}
Similarly to the scalar case the generalization of the central charge is always decreasing, starting from the UV value
\begin{equation}
 \begin{split}
 C(z)&= \frac{d_\gamma}{2} +{\cal O}\left(\frac{m^2}{q^2}\right) \qquad {\rm for} \quad q^2 \gg m^2\,.
 \end{split}
\end{equation}
This agrees with the fact that $c=\frac{1}{2}$ is the expected central charge of a single fermionic degree of freedom in $D=2$.

\subsection{Proca field in two dimensions}\label{sect:proca2d}

Finally we report all the terms needed for the Proca field trace appearing in \eqref{eq:functional-traces} in $D=2$
to be used in conjunction with \eqref{eq:effective-action-2d}.
The local part of the effective action is
\begin{equation}
 \begin{split}
 \Gamma_{\rm loc}[g] &=
 \frac{1}{4\pi}\int {\rm d}^2 x \sqrt{g}\, \Bigl\{
 \left(\frac{1}{\bar{\epsilon}}+\frac{1}{2}\right)m^2
  +\frac{5}{6}\frac{1}{\bar{\epsilon}} R
 \Bigr\}\,.
 \end{split}
\end{equation}
The non-local part of \eqref{eq:effective-action-2d} is captured by the functions
\begin{equation}
 \begin{split}
 \cB(z)
 = \frac{1}{36}+\frac{3Y}{4}+\frac{Y}{3a^2}\,,
 &\qquad
 \cC(z)
 =
 -\frac{1}{2}+3Y- \frac{6Y}{a^2}+\frac{3}{8}(1-Y)a^2\,.
 \end{split}
\end{equation}
The non-local beta function related to the running of the Newton's constant is
\begin{equation}
 \begin{split}
 \beta_{b_1}
 &= \frac{1}{2\pi}\left\{-\frac{1}{24}-\frac{Y}{4} - \frac{Y}{2a^2}-\frac{3}{32}(1-Y)a^2\right\}\,,
 \end{split}
\end{equation}
and it has the limits
\begin{equation}
 \begin{split}
 \beta_{b_1} &= \begin{cases}
            -\frac{5}{24\pi} +{\cal O}\left(\frac{m^2}{q^2}\right) & \qquad {\rm for} \quad q^2 \gg m^2 \,; \\
           -\frac{1}{30\pi} \frac{q^2}{m^2}+{\cal O}\left(\frac{q^2}{m^2}\right)^{\frac{3}{2}} & \qquad {\rm for} \quad q^2 \ll m^2\,.
           \end{cases}
 \end{split}
\end{equation}
The Proca field is not conformally coupled neither for non-zero mass, nor in the limit $m\to 0$.
In fact, the conformally coupled ``equivalent'' of the Proca field is a Maxwell field, but we have established in section \ref{sect:effective-action} that such limit is discontinuous.
Nevertheless in the ultraviolet
\begin{equation}
 \begin{split}
C(\infty)&= 1 +{\cal O}\left(\frac{m^2}{q^2}\right) \qquad {\rm for} \quad q^2 \gg m^2\,,
 \end{split}
\end{equation}
which correctly counts the number of degrees of freedom for a
Proca field in $D=2$ (two degrees of freedom of a vector minus one from the ghost scalar).

\section{Renormalized action in four dimensions}\label{sect:nonlocal-four}

In four dimensions the regularized effective action is much more complicate than the one shown in section \ref{sect:nonlocal-two}.
As general template for its parametrization we define
\begin{equation}
 \begin{split} \label{eq:effective-action-full}
\Ga[g] &=
\Ga_{\rm loc}[g]
 + \frac{m^2}{2(4\pi)^2}\int {\rm d}^4 x \sqrt{g}\,
 B(z) R
 + \frac{1}{2(4\pi)^2}\int {\rm d}^4 x \sqrt{g}\, \Bigl\{
 C^{\mu\nu\alpha\beta} \, C_1(z) \, C_{\mu\nu\alpha\beta}
 + R \, C_2(z) \, R
 \Bigr\}\,,
 \end{split}
\end{equation}
in which we used the four-dimensional Weyl tensor $C_{\mu\nu\rho\theta}$. In our settings the non-local functions $C_1(z)$ and $C_2(z)$
are four-dimensional generalizations of $\cC(z)$ and therefore we could speculate on their relations with the $a$- and $c$-charges
that appear in four-dimensional generalizations of Zamolodchikov's analysis \cite{Jack:1990eb} through local RG \cite{Osborn:1991gm}.
It would be intriguing to establish a connection with the functional formalism of \cite{Codello:2015ana} but we do not dive further in this direction.

The heat kernel terms that require renormalization are those with zero, one and two curvatures, corresponding to poles coming from
the integration of
$s^{-d/2}$, $s^{-d/2+1}$ and $s^{-d/2+2}$.
All the poles are local, which means that they are contained in
$\Ga_{\rm loc}[g]$ and can be renormalized by introducing the counterterms.
The renormalized local action is
\begin{equation}
 \begin{split} \label{eq:effective-action-local-renormalized}
 S_{\rm ren}[g]
 &= \int {\rm d}^4 x \sqrt{g}\,\left\{b_0 + b_1 R + a_1 C^2 +a_2 {\cal E}_4 
 +a_3 \Box R + a_4 R^2\right\}\,,
 \end{split}
\end{equation}
in which ${\cal E}_4$ is the operator associated to the Euler's characteristic, which is the 
Gauss-Bonnet topological term in $d=4$.
Our non-local heat kernel of appendix \ref{sect:heat-kernel} is valid for asymptotically flat spacetimes,
which has the unfortunate consequence of setting  ${\cal E}_4=0$, but we can study every other
term flawlessly \cite{bavi90}.
The couplings of \eqref{eq:effective-action-local-renormalized} include the cosmological constant $\Lambda$ and 
the Newton's constant $G$ through the relations $b_0=2\Lambda G^{-1}$
and $b_1=-G^{-1}$. In general, we denote beta functions in the minimal subtraction scheme
as $\beta^{\overline{\rm MS}}_g$ in which $g$ is any of the couplings appearing in \eqref{eq:effective-action-local-renormalized}.

Comparing \eqref{eq:effective-action-local-renormalized} with \eqref{eq:effective-action-full} we can straightforwardly
define the non-local renormalization group beta function for two of the quadratic couplings
\begin{equation}\label{eq:definitions-running-1}
 \begin{split}
 \beta_{a_1} =
  \frac{z}{(4\pi)^2} C'_1(z)\,,\qquad
 \beta_{a_4} =  \frac{z}{(4\pi)^2} C'_2(z)\,,
 \end{split}
\end{equation}
and these definitions coincide with the ones made in \cite{apco,fervi}.
In contrast to the two-dimensional case, it is much less clear how to attribute the running of the function $B(z)$
because both $R$ and $\Box R$ require counterterms. We discuss some implications of this point in section \ref{sect:uv-structure}.
To handle the problem
we define a master ``beta function'' for the couplings that are linear in the scalar curvature
\begin{equation}\label{eq:psi-def}
 \begin{split}
 \Psi ~ = ~ \frac{1}{(4\pi)^2}\, z\,\partial_z \Big[\frac{B(z)}{z}\Big]\,.
 \end{split}
\end{equation}
The function $\Psi$ includes the non-local running of both couplings $a_3$ and $b_1$,
which can be seen from the general property
\begin{equation}\label{eq:psi}
 \begin{split}
  \Psi &=  \begin{cases}
            - \beta^{\overline{\rm MS}}_{a_3} & \qquad {\rm for} \quad q^2 \gg m^2 \\
            \frac{m^2}{q^2} ~ \beta^{\overline{\rm MS}}_{b_1} & \qquad {\rm for} \quad q^2 \ll m^2
           \end{cases}
 \end{split}
\end{equation}
that we observe for all the matter species that we considered. The function $\Psi$ ``mutates'' from the ultraviolet to the infrared
giving the universal $\overline{\rm MS}$ contributions of the running of both $a_3$ and $b_1$.
Following the discussion of section \ref{sect:uv-structure} we define the non-local beta functions by clearing the asymptotic behaviors
\begin{equation}
 \begin{split} 
 \label{eq:definitions-running}
 \beta_{a_3} ~ = ~ -\frac{1}{(4\pi)^2} \,z\,\partial_z 
 \Big[\frac{B(z)-B(0)}{z}\Big], 
 \qquad \beta_{b_1} ~ = ~ \frac{m^2}{(4\pi)^2} \,z\,
 \partial_z\big[B(z)-B_\infty(z)\big].
 \end{split}
\end{equation}
In order to preserve the elegance of the form-factors and of the beta functions expressed only in terms of the dimensionless variables $a$ and $Y$,
instead of subtracting the leading logarithm at infinity we subtract
\begin{equation}
 \begin{split}
  a(1-Y)\simeq \ln (z) \,,
 \end{split}
\end{equation}
which is shown to be valid for $z\gg 1$ using the definitions 
\eqref{eq:dimensionless-operators}.

Using the above definitions \eqref{eq:definitions-running-1} and \eqref{eq:definitions-running}, each separate beta function coincides with its $\overline{\rm MS}$ 
counterpart in the ultraviolet
\begin{equation}
 \begin{split}
 \beta_g 
 &= \beta^{\overline{\rm MS}}_g 
 + {\cal O}\left(\frac{m^2}{q^2}\right) \qquad {\rm for }\qquad q^2\gg m^2,
 \end{split}
\end{equation}
in which $g$ is any of the couplings of \eqref{eq:effective-action-full} (with the possible exception of $a_2$ which is not present in asymptotically flat spacetimes).
Furthermore, in the infrared the running of each coupling is slowed down by a quadratic factor of the energy
\begin{equation}
 \begin{split}
 \beta_g 
 &=  {\cal O}\left(\frac{q^2}{m^2}\right) \qquad {\rm for }\qquad q^2 \ll m^2\,,
 \end{split}
\end{equation}
which is a practical evidence of the Appelquist-Carazzone theorem in a four-dimensional space.

\subsection{Non-minimally coupled scalar field in four dimensions}\label{sect:scalar4d}

The effective action of the non-minimally coupled scalar field can be obtained specifying the endomorphism $E=\xi R$
in the non-local heat kernel expansion and then performing the integration in $s$. We give all the results using the template \eqref{eq:effective-action-full}.
We find the local contributions of the regularized action to be
\begin{multline}
 \Ga_{\rm loc}[g] =
 \frac{1}{2(4\pi)^2} \int {\rm d}^4 x\sqrt{g} \, \Bigl\{
 -m^4\Bigl(\frac{1}{\bar{\epsilon}}+\frac{3}{4}\Bigr)
 - 2m^2\Bigl( \xi-\frac{1}{6} \Bigr)\frac{1}{\bar{\epsilon}} R
  \\
 +\frac{1}{3} \Bigl( \xi-\frac{1}{5} \Bigr)\frac{1}{\bar{\epsilon}} \Box R
 -\frac{1}{60\bar{\epsilon}} C_{\mu\nu\rho\theta} C^{\mu\nu\rho\theta}
 -\Bigl( \xi-\frac{1}{6} \Bigr)^2 \frac{1}{\bar{\epsilon}} R^2
 \Bigr\}\,.
\end{multline}
The minimal subtraction of the divergences of local contributions induces the following $\overline{\rm MS}$ running
\begin{equation}\label{eq:beta-functions-scalar-perturbative}
\begin{array}{lll}
 \beta_{b_0}^{\overline{\rm MS}} = \frac{1}{(4\pi)^2} \frac{m^4}{2} \,,
 &  \quad
 \beta_{b_1}^{\overline{\rm MS}} = \frac{1}{(4\pi)^2} m^2 \xibar \,,
 & \quad \\
 \beta_{a_3}^{\overline{\rm MS}} = - \frac{1}{(4\pi)^2} \frac{1}{6} \left(\xi-\frac{1}{5}\right)\,,
 & \quad
 \beta_{a_1}^{\overline{\rm MS}} = \frac{1}{(4\pi)^2}\frac{1}{120}\,,
 & \quad
 \beta_{a_4}^{\overline{\rm MS}} = \frac{1}{(4\pi)^2} \frac{1}{2} \xibar^2\,,
\end{array}
\end{equation}
which agree with \cite{bro-cass,Jack:1983sk,Martini:2018ska} in the overlapping region of validity.
The non-local part of the effective action includes the following form-factors
\begin{equation}
 \begin{split}
 \frac{B(z)}{z} &=
-\frac{4 Y}{15 a^4}+\frac{Y}{9 a^2}-\frac{1}{45 a^2}+\frac{4}{675}+\xibar \left(-\frac{4 Y}{3 a^2}-\frac{1}{a^2}+\frac{5}{36}\right)\,,
 \\
 C_1(z) &= -\frac{1}{300}-\frac{1}{45a^2}-\frac{4Y}{15a^4} \,,
 \\
 C_2(z) &=
 -\frac{Y}{144}+\frac{7}{2160}-\frac{Y}{9 a^4}+\frac{Y}{18 a^2}-\frac{1}{108 a^2}
 +\xibar \left(-\frac{2 Y}{3 a^2}+\frac{Y}{6}-\frac{1}{18}\right)-Y \xibar^2 \,.
 \end{split}
\end{equation}
Using our definitions \eqref{eq:definitions-running-1} and \eqref{eq:definitions-running} the non-local beta functions are
\begin{equation}
 \begin{split}
 \beta_{b_1} &=\frac{z}{(4\pi)^2}\Bigl\{
 \frac{2 Y}{5 a^4}-\frac{2 Y}{9 a^2}
 + \frac{1}{30 a^2}-\frac{aY}{180}
 +\frac{a}{120}+\frac{Y}{24}-\frac{1}{40}
 +
  \xibar \left(\frac{2 Y}{3 a^2}+\frac{a Y}{6}-\frac{a}{4}
  -\frac{Y}{2}+\frac{1}{2}\right)
  \Bigr\}\,,
  \\
 \beta_{a_3} &= \frac{1}{(4\pi)^2}
 \Bigl\{
 -\frac{2 Y}{3 a^4}
 +\frac{Y}{3 a^2}-\frac{1}{18 a^2}
 -\frac{Y}{24}+\frac{7}{360}
 +
 \xibar
 \left(-\frac{2 Y}{a^2}+\frac{Y}{2}-\frac{1}{6}\right)
 \Bigr\}\,, \\
 \beta_{a_1} &= \frac{1}{(4\pi)^2}\Bigl\{
  -\frac{1}{180}+\frac{1}{18a^2}+\frac{2Y}{3a^4}-\frac{Y}{6a^2}
  \Bigr\}\,,
  \\
 \beta_{a_4} &= \frac{1}{(4\pi)^2}\Bigl\{
  \frac{5 Y}{18 a^4}-\frac{a^2 Y}{1152}-\frac{11 Y}{72 a^2}+\frac{a^2}{1152}+\frac{5}{216 a^2}+\frac{7 Y}{288}-\frac{1}{108}
  \\&
  \quad +\xibar \left(\frac{a^2 Y}{48}+\frac{Y}{a^2}-\frac{a^2}{48}-\frac{Y}{3}+\frac{1}{12}\right)
  +\xibar^2 \left(-\frac{a^2 Y}{8}+\frac{a^2}{8}+\frac{Y}{2}\right)
  \Bigr\}\,.
 \end{split}
\end{equation}
The effects of the Appelquist-Carazzone for $\beta_{a_1}$ and $\beta_{a_4}$ have been observed in \cite{apco,fervi},
and for $\beta_{b_1}$ and $\beta_{a_3}$ in \cite{Franchino-Vinas:2018gzr}. We report the latter two because they are related to the Newton's constant through $b_1=-G^{-1}$.
The non-local beta function of the coupling $b_1$ in units of the mass has the two limits
\begin{equation}
\begin{split}
\frac{\beta_{b_1}}{m^2} &=  \begin{cases}
\frac{1}{(4\pi)^2}\xibar + \frac{1}{(4\pi)^2}
\left\{\left(\frac{3}{5}-\xi\right)
-\xi \ln\left(\frac{q^2}{m^2}\right)\right\} \frac{m^2}{q^2}
+{\cal O}\left(\frac{m^2}{q^2}\right)^{2}
&
\quad {\rm for} \quad q^2 \gg m^2\,,
\\
\frac{1}{(4\pi)^2}\left(\frac{4}{9}\xi-\frac{77}{900}\right)
\frac{q^2}{m^2} +{\cal O}\left(\frac{q^2}{m^2}\right)^{\frac{3}{2}}
&
\quad {\rm for} \quad q^2 \ll m^2\,;
\end{cases}
 \end{split}
\end{equation}
while the one of $a_3$ is
\begin{equation}
\begin{split}
\beta_{a_3} &= \begin{cases}
-\frac{1}{6(4\pi)^2}\left(\xi-\frac{1}{5}\right)
+ \frac{1}{(4\pi)^2}  \left\{ \frac{5}{18}-2\xi
+ \xibar \ln\left(\frac{q^2}{m^2}\right)\right\} \frac{m^2}{q^2}
            +{\cal O}\left(\frac{m^2}{q^2}\right)^{2}
& \,\,\, {\rm for} \,\,\, q^2 \gg m^2\,,
\\
\frac{1}{(4\pi)^2} \frac{1}{840}
\left(3-14\xi\right)\frac{q^2}{m^2}
+{\cal O}\left(\frac{q^2}{m^2}\right)^2
& \,\,\, {\rm for} \,\,\, q^2 \ll m^2\,.
\end{cases}
 \end{split}
\end{equation}
These expressions show a standard quadratic decoupling in the IR, 
exactly as for QED \cite{AC} and the fourth 
derivative gravitational terms \cite{apco,fervi}.

\subsection{Dirac field in four dimensions}\label{sect:dirac4d}

The effective action of the minimally coupled Dirac fields requires the specification of the endomorphism $E=R/4$.
The final result is proportional to the dimension $d_\gamma$ of the Clifford algebra and hence to
the number of spinor components. We do not set $d_\gamma=4$, but choose instead to leave it arbitrary
so that the formulas can be generalized to other spinor species easily.
We find the local regularized action to be
\begin{equation}
 \begin{split}
 \Ga_{\rm loc}[g] &=\frac{d_\gamma}{2(4\pi)^2} \int {\rm d}^4 x\sqrt{g} \, \Bigl\{
 m^4\Bigl(\frac{1}{\bar{\epsilon}}+\frac{3}{4}\Bigr)
 +\frac{m^2}{6\bar{\epsilon}} R -\frac{1}{60\bar{\epsilon}} \Box R
 -\frac{1}{40\bar{\epsilon}} C_{\mu\nu\rho\theta} C^{\mu\nu\rho\theta}
 \Bigr\}\,.
 \end{split}
\end{equation}
The minimal subtraction of the $1/\bar{\epsilon}$ divergences induces the following $\overline{\rm MS}$ beta functions
\begin{equation}
\begin{array}{lll}
 \beta_{b_0}^{\overline{\rm MS}} = -\frac{d_\gamma}{(4\pi)^2} \frac{m^4}{2} \,,
 &  \quad
 \beta_{b_1}^{\overline{\rm MS}} = -\frac{d_\gamma}{(4\pi)^2} \frac{m^2}{12} \,,
 & \quad \\
 \beta_{a_3}^{\overline{\rm MS}} = \frac{d_\gamma}{(4\pi)^2} \frac{1}{120} \,,
 & \quad
 \beta_{a_1}^{\overline{\rm MS}} = \frac{d_\gamma}{(4\pi)^2}\frac{1}{80} \,,
 & \quad
 \beta_{a_4}^{\overline{\rm MS}} = 0\,.
\end{array}
\end{equation}
The non-local part of the effective action includes the following form-factors
\begin{equation}
 \begin{split}
 \frac{B(z)}{z} &=
 d_\gamma\Bigl\{-\frac{7}{400}+\frac{19}{180a^2}+\frac{4Y}{15a^4} \Bigr\}\,,
 \\
 C_1(z) &= d_\gamma\Bigl\{-\frac{19}{1800}+\frac{1}{45a^2}+\frac{4Y}{15a^4}-\frac{Y}{6a^2}\Bigr\} \,,
 \\
 C_2(z) &= d_\gamma\Bigl\{-\frac{1}{1080}+\frac{1}{108a^2}+\frac{Y}{9a^4}-\frac{Y}{36a^2}\Bigr\} \,.
 \end{split}
\end{equation}
The non-local beta functions are
\begin{equation}
 \begin{split}
 \beta_{b_1} &= \frac{d_\gamma z}{(4\pi)^2}\Bigl\{-\frac{2 Y}{5 a^4}+\frac{Y}{6 a^2}-\frac{1}{30 a^2}-\frac{a Y}{120}+\frac{a}{80}-\frac{1}{60} \Bigr\}\,,
  \\
  \beta_{a_3} &= \frac{d_\gamma}{(4\pi)^2} \Bigl\{ \frac{2 Y}{3 a^4}-\frac{Y}{6 a^2}+\frac{1}{18 a^2}-\frac{1}{180}\Bigr\}\,,
  \\
 \beta_{a_1} &= \frac{d_\gamma}{(4\pi)^2}\Bigl\{-\frac{2 Y}{3 a^4}+\frac{5 Y}{12 a^2}-\frac{1}{18 a^2}-\frac{Y}{16}+\frac{19}{720} \Bigr\}\,,
  \\
 \beta_{a_4} &= \frac{d_\gamma}{(4\pi)^2}\Bigl\{-\frac{5 Y}{18 a^4}+\frac{Y}{9 a^2}-\frac{5}{216 a^2}-\frac{Y}{96}+\frac{5}{864} \Bigr\}\,.
 \end{split}
\end{equation}
Likewise in the scalar case the non-local beta functions of $b_1$ and $a_3$ 
have two limits
\begin{equation}
\begin{split}
\frac{\beta_{b_1}}{m^2}
&=  \begin{cases}
            -\frac{d_\gamma}{(4\pi)^2}\frac{1}{12}
            -\frac{d_\gamma}{(4\pi)^2}\left[\frac{7}{20}
            -\frac{1}{4}\ln\left(\frac{q^2}{m^2}\right)\right]\frac{m^2}{q^2} +{\cal O}\left(\frac{m^2}{q^2}\right)^{{2}}
& \qquad {\rm for} \quad q^2 \gg m^2\,;
\\
            -\frac{d_\gamma}{(4\pi)^2}\frac{23}{900} \frac{q^2}{m^2} +{\cal O}\left(\frac{q^2}{m^2}\right)^{\frac{3}{2}}
& \qquad {\rm for} \quad q^2 \ll m^2\,.
           \end{cases}\\
  \beta_{a_3} &= \begin{cases}
            \frac{d_\gamma}{(4\pi)^2} \frac{1}{120} + \frac{d_\gamma}{(4\pi)^2}\left\{\frac{2}{9}
            -\frac{1}{12}\ln\left(\frac{q^2}{m^2}\right)\right\}\frac{m^2}{q^2}
            +{\cal O}\left(\frac{m^2}{q^2}\right)^{{2}}
            & \qquad {\rm for} \quad q^2 \gg m^2 \,;
            \\
            \frac{d_\gamma}{(4\pi)^2} \frac{1}{1680} \frac{q^2}{m^2}
            +{\cal O}\left(\frac{q^2}{m^2}\right)^2
            & \qquad {\rm for} \quad q^2 \ll m^2\,.
           \end{cases}
 \end{split}
\end{equation}
As in the previous section there is the standard quadratic decoupling in the IR.

\subsection{Proca field in four dimensions}\label{sect:proca4d}

The integration of the minimally coupled Proca field exhibits the local regularized action
\begin{equation}
 \begin{split}
 \Ga_{\rm loc}[g] &=
 \frac{1}{2(4\pi)^2} \int {\rm d}^4 x\sqrt{g} \, \Bigl\{
 -m^4\Bigl(\frac{3}{\bar{\epsilon}}+\frac{9}{4}\Bigr)
 -\frac{m^2}{\bar{\epsilon}} R +\frac{2}{15\bar{\epsilon}} \Box R
 -\frac{13}{60\bar{\epsilon}} C_{\mu\nu\rho\theta} C^{\mu\nu\rho\theta}
 -\frac{1}{36} R^2
 \Bigr\}\,.
 \end{split}
\end{equation}
The minimal subtraction of the $1/\bar{\epsilon}$ poles induces the 
following $\overline{\rm MS}$ beta functions
\begin{equation}
\begin{array}{lll}
 \beta_{b_0}^{\overline{\rm MS}} = \frac{1}{(4\pi)^2} \frac{3m^4}{2} \,,
 &  \quad
 \beta_{b_1}^{\overline{\rm MS}} = \frac{1}{(4\pi)^2} \frac{m^2}{2} \,,
 & \quad \\
 \beta_{a_3}^{\overline{\rm MS}} = -\frac{1}{(4\pi)^2} \frac{1}{15} \,,
 & \quad
 \beta_{a_1}^{\overline{\rm MS}} = \frac{1}{(4\pi)^2}\frac{13}{120} \,,
 & \quad
 \beta_{a_4}^{\overline{\rm MS}} = \frac{1}{(4\pi)^2}\frac{1}{72}\,.
\end{array}
\end{equation}
The non-local part of the effective action includes the following form-factors
\begin{equation}
 \begin{split}
 \frac{B(z)}{z} &= \frac{157}{1800} -\frac{17}{30 a^2} -\frac{4 Y}{5 a^4}-\frac{Y}{3 a^2} \,,
 \\
 C_1(z) &= \frac{91}{900} -\frac{1}{15 a^2}-\frac{Y}{2} -\frac{4 Y}{5 a^4}+\frac{4 Y}{3 a^2} \,,
 \\
 C_2(z) &= \frac{1}{2160} -\frac{1}{36 a^2} -\frac{Y}{3 a^4} -\frac{Y}{48} +\frac{Y}{18 a^2} \,.
 \end{split}
\end{equation}
The non-local beta functions are easily derived
\begin{equation}
 \begin{split}
 \beta_{b_1} &= \frac{z}{(4\pi)^2} \left\{ \frac{6 Y}{5 a^4}-\frac{Y}{3 a^2}+\frac{1}{10 a^2}+\frac{a Y}{15}-\frac{a}{10}-\frac{Y}{8}+\frac{7 }{40} \right\}\,,
  \\
  \beta_{a_3} &= \frac{1}{(4\pi)^2} \Bigl\{ -\frac{2 Y}{a^4}-\frac{1}{6 a^2}+\frac{Y}{8}-\frac{1}{40} \Bigr\}\,,
  \\
 \beta_{a_1} &= \frac{1}{(4\pi)^2}\Bigl\{\frac{2 Y}{a^4}-\frac{a^2 Y}{16}-\frac{5 Y}{2 a^2}+\frac{a^2}{16}+\frac{1}{6 a^2}+\frac{3 Y}{4}-\frac{11}{60}\Bigr\}\,,
  \\
 \beta_{a_4} &= \frac{1}{(4\pi)^2}\Bigl\{\frac{5 Y}{6 a^4}-\frac{a^2 Y}{384}-\frac{7 Y}{24 a^2}+\frac{a^2}{384}+\frac{5}{72 a^2}+\frac{Y}{32}-\frac{1}{72} \Bigr\}\,.
 \end{split}
\end{equation}
The beta functions of $b_1$ and $a_3$ have the two limits
\begin{equation}
 \begin{split}
  \frac{\beta_{b_1}}{m^2} &=  \begin{cases}
            \frac{1}{(4\pi)^2}\frac{1}{2}+\frac{1}{(4\pi)^2}\left(\frac{4}{5}
            -\ln\left(\frac{q^2}{m^2}\right)\right)\frac{m^2}{q^2} +{\cal O}\left(\frac{m^2}{q^2}\right)^{2}
  & \qquad {\rm for} \quad q^2 \gg m^2\,;
            \\
            \frac{1}{(4\pi)^2}\frac{169}{900} \frac{q^2}{m^2}
            +{\cal O}\left(\frac{q^2}{m^2}\right)^{\frac{3}{2}}
  & \qquad {\rm for} \quad q^2 \ll m^2\,,
           \end{cases}\\
  \beta_{a_3} &= \begin{cases}
            -\frac{1}{(4\pi)^2} \frac{1}{15}
            - \frac{1}{(4\pi)^2}\left\{\frac{7}{6}
            -\frac{1}{2}\ln\left(\frac{q^2}{m^2}\right)\right\}\frac{m^2}{q^2}
            +{\cal O}\left(\frac{m^2}{q^2}\right)^{2}
            & \qquad {\rm for} \quad q^2 \gg m^2 \,;
            \\
            -\frac{1}{(4\pi)^2} \frac{1}{168} \frac{q^2}{m^2}
            +{\cal O}\left(\frac{q^2}{m^2}\right)^2 & \qquad {\rm for}
            \quad q^2 \ll m^2\,.
           \end{cases}
 \end{split}
\end{equation}
We can observe that also for the Proca field there is a quadratic decoupling. 

\section{Comments on the UV structure of the effective action}\label{sect:uv-structure}

The local and non-local contributions to the effective action \eqref{eq:effective-action-full} are not fully independent,
but rather display some important relations which underline the properties described in Sect.~\ref{sect:nonlocal-four}.
We concentrate here on the running of a generic operator $O[g]$ on which a form-factor $B_O(z)$ acts,
while keeping in mind that the explicit example would be to take $R$ as the operator and $B(z)$ as the corresponding form-factor.
For small mass $m\sim 0$ we expect on general grounds that the regularized vacuum action is always of the form
\begin{equation}
 \begin{split}\label{eq:gamma-local-div}
 \Ga[g] &\supset
 - \frac{b _O}{(4\pi)^2\bar{\epsilon}}  \int {\rm d}^4x~O[g]
 + \frac{1}{2(4\pi)^2}\int {\rm d}^4x~ B_O(z)~O[g]
 \\
 &
 = -\frac{b _O}{2(4\pi)^2}\int {\rm d}^4x \Bigl[\frac{2}{\bar{\epsilon}} - \ln\left(\D/m^2\right)\Bigr]O[g] + \dots
 \end{split}
\end{equation}
which can be proven coupling $O[g]$ to the path integral as a scalar composite operator.
The dots hide subleading contributions in the mass and $b_O$ is a unique coefficient determined by the renormalization of the operator itself.
The above relation underlines the explicit connection between the coefficient of the $1/\bar{\epsilon}$ pole and the leading
ultraviolet logarithmic behavior of the form-factor \cite{El-Menoufi:2015cqw,Donoghue:2015nba}.

The subtraction of the pole requires the introduction of the renormalized coupling $g_O$
\begin{equation}
 \begin{split}
 S_{\rm ren}[g] \supset \int g_O ~ O[g]\,,
 \end{split}
\end{equation}
which in the $\overline{\rm MS}$ scheme will have the beta function
\begin{equation}
 \begin{split}
 \beta^{\overline{\rm MS}}_{g_O} &= \frac{b _O}{(4\pi)^2}\,.
 \end{split}
\end{equation}
Following our discussion of section \ref{sect:nonlocal-four} we find that
if we subtract the divergence at the momentum scale $q^2$ coming from the Fourier transform of the form-factor
we get a non-local beta function
\begin{equation}
 \begin{split}
 \beta_{g_O} &= \frac{z}{(4\pi)^2} B'_{g_O}(z)\,.
 \end{split}
\end{equation}
Using \eqref{eq:gamma-local-div} it is easy to see that in the ultraviolet limit $z\gg 1$
\begin{equation}
 \begin{split}
  B(z)&= b _O \ln\left(z\right) + \dots \,,
 \end{split}
\end{equation}
from which one can infer in general that the ultraviolet limit of the non-local beta function coincides with the $\overline{\rm MS}$ result
\begin{equation}\label{eq:running-of-gO}
 \begin{split}
  \beta_{g_O}&=\beta^{\overline{\rm MS}}_{g_O} +\dots \qquad{\rm for}\quad z\gg 1\,.
 \end{split}
\end{equation}

It might not be clear at a first glance, but in the above discussion we are implicitly assuming that the operator $O[g]$
is kept fixed upon actions of the renormalization group operator $q\partial q=2z\partial_z$.
Suppose instead that the operator $O[g]$ is actually a total derivative of the form
\begin{equation}
 \begin{split}
  O[g] = \Box \, O'[g] = -\Delta_g O'[g] \,,
 \end{split}
\end{equation}
in which we introduce another operator $O'[g]$ to be renormalized with a coupling $g_{O'}$ and a local term $g_{O'}\int\Box O'[g]$.
If we act with $q\partial_q$ and keep $O'[g]$ fixed instead of $O[g]$ we get
\begin{equation}
 \begin{split}
  \beta_{g_{O'}}O'[g] \propto -\frac{1}{2(4\pi)^2} q\partial_q\left( B_O(z)O[g]\right) = -z\beta _O O'[g] - \frac{1}{(4\pi)^2} z B_O(z) O'[g] \,.
 \end{split}
\end{equation}
Obviously we find an additional scaling term proportional to the form-factor $B_O(z)$ itself. The definitions \eqref{eq:definitions-running}
take care of this additional scaling by switching the units of $B(z)$ before applying the derivative with respect to the scale.
In the general example of this appendix we would follow this strategy by defining
\begin{equation}\label{eq:running-of-gOprime}
 \begin{split}
  \beta_{g_{O'}} = -\frac{1}{(4\pi)^2} z\partial_z\left(\frac{B_O(z)}{z}\right)\,,
 \end{split}
\end{equation}
for the running of the total derivative coupling.

The definitions \eqref{eq:running-of-gO} and \eqref{eq:running-of-gOprime} now ensure the correct scaling behavior of the running,
but are still sensitive to some problems, as shown in practice by \eqref{eq:psi}. These problems are related to the fact
that some terms that should be attributed in the UV/IR limits of either coupling's running appear in the other coupling's running.
For example, our mass-dependent running of $\Box R$ dominates $\Psi$ in the ultraviolet because $\Box\sim -q^2$ grows unbounded,
while the same happens in the infrared for $R$.
In \eqref{eq:definitions-running} of the main text we have adopted the convention of subtracting the asymptotic (clearly attributable) behavior
of either coupling to the definition of the running of the other coupling as follows
\begin{equation}\label{eq:running-of-g0-gOprime-sub}
 \begin{split}
  \beta_{g_{O}} = -\frac{1}{(4\pi)^2} z\partial_z\left(B_O(z)-B_{\infty,O}(z)\right)\,, \qquad\beta_{g_{O'}} = -\frac{1}{(4\pi)^2} z\partial_z\left(\frac{B_O(z)-B_O(0)}{z}\right)\,,
 \end{split}
\end{equation}
in which $B_{\infty,O}(z)$ is the asymptotic behavior of $B_O(z)$ at $z=\infty$ (see the discussion of section \ref{sect:nonlocal-four} for the practical application).
These definitions ensure that the dimensional $\overline{\rm MS}$ beta functions of both couplings are reproduced in the UV if both couplings require counterterms,
and have the important property of agreeing with the predictions of the Appelquist-Carazzone theorem in the infrared.

\section{Scheme dependence and quantum gravity}
\label{sect:scheme}

In this section we speculate on possible uses of the framework described in sections \ref{sect:nonlocal-two} and \ref{sect:nonlocal-four} to the context of quantum gravity
and, more specifically, of asymptotically safe gravity \cite{Reuter:1996cp,books}.
We begin by recalling that the asymptotic safety conjecture suggests that the four-dimensional quantum theory of metric gravity might be asymptotically safe.
An asymptotically safe theory is one in which the ultraviolet is controlled by a non-trivial fixed point of the renormalization group with a finite number of UV relevant directions.
Therefore the first and most important point to validate the asymptotic safety conjecture is thus to show that the gravitational couplings,
in particular the Newton's constant, have a non-trivial fixed point in their renormalization group flow.

On general grounds, the RG of quantum gravity is induced by the integration of gravitons and all other fields,
with the latter including both all matter flavors and types and gauge fields. Certainly in this review we have not considered gauge nor graviton fields, 
but we can still capture some information of a presumed fixed point. If for example
quantum gravity is coupled to a large number of minimally coupled scalar fields, $n_{\rm s}\gg 1$,
then we can assume with reasonable certainty that fluctuations of the scalar fields will dominate the running in the large-$n_{\rm s}$
expansion and we could promote \eqref{eq:beta-functions-scalar-perturbative} using $b_1=-G^{-1}$ and $\xi=0$ to obtain the beta function $\beta_G$ \cite{Martini:2018ska,Codello:2011js}
without having to deal with gauge-fixing and ghosts \cite{Eichhorn:2009ah,Groh:2010ta}.

One point of criticism of the use of $\beta_G$ for making physical predictions is that the running of Newton's constant is
strongly dependent on the scheme in which it is computed. If we use dimensional regularization and assume that $n_{\rm s}$ is large,
we have the counterterm relation
\begin{equation}
 \begin{split}
 -\frac{1}{G_0}
 = \mu^{\epsilon}\left( -\frac{1}{G} - n_{\rm s} \,\frac{ m^2}{6(4\pi)^2\epsilon} \right)\,;
 \end{split}
\end{equation}
if instead we use any scheme involving a cutoff $\Lambda$
\begin{equation}\label{eq:cc-scheme}
 \begin{split}
 -\frac{1}{G_0}
 =
 -\frac{1}{G} + A_{\rm sch} \, \Lambda^2
 - n_{\rm s}\,\frac{m^2}{6(4\pi)^2}\log\Lambda\,,
 \end{split}
\end{equation}
in which we introduced the constant $A_{\rm sch}$ that depends on the specific details of the scheme.
We can see that the coefficient of the dimensional pole of the $\overline{\rm MS}$ subtraction is universal: it survives the change of scheme
and it multiplies the logarithm in the massive scheme.
This is of course a well-known relation of quantum field theory.

The vast body of literature dedicated to the conjecture points to the fact that the existence of the fixed point
hinges on the inclusion of the scheme dependent part, but this is often reason of mistrust because the
quantities that are computed using $A_{\rm sch}$ depend on the scheme in very complicate ways, especially 
if considered beyond the limitations of perturbation theory.
In short there are two very polarized points of view on the credibility of results based on \eqref{eq:cc-scheme}
which seem impossible to make agree conceptually.
Ideally, in order to find common ground between the points of view,
one would like to have a relation almost identical to \eqref{eq:cc-scheme}, but in which $\Lambda$
is replaced by some scale $q^2$ which has physical significance, meaning that it is related to some momentum of a given magnitude.
Our definition of renormalization group as given in \eqref{eq:psi-def} and \eqref{eq:definitions-running}
does something very close, in that $q^2$ is a momentum variable of a form-factor which
could in principle be related to some gravitational observable.

The function $B(z)$ could thus work as a scale dependent Newton's constant and $\Psi(z)$ as its beta function
in the usual sense required by asymptotic safety,
yet they could maintain some physical meaning thanks to the momentum scale $q^2$. From this point of view the scheme dependence
of \eqref{eq:cc-scheme} could be replaced by the dependence of the renormalization condition,
hence on the appropriate observable that incorporates $B(z)$ and the scale $q^2$.
This idea is certainly \emph{very} speculative, but it becomes worth considering after identifying an interesting conclusion:
we have observed in \eqref{eq:psi} that $\Psi(z)$ always has two limits: in the infrared it reproduces the universal running
of the Newton's constant, while in the ultraviolet it reproduces the universal running of the coupling of $\Box R$.
This fact might be suggesting that in determining the ultraviolet nature of quantum gravity the operator $\Box R$
plays the role commonly associated to $R$.
We hope that our results might offer some inspiration for further developments
in the direction of a more formal proof of the asymptotic safety conjecture.

\section{Conclusions}
\label{sect:conclusions}

We reviewed the covariant computation of the non-local form-factors of the metric-dependent effective action
which integrates the effects of several massive matter fields over two- and four-dimensional metric Euclidean spacetime.
We established a connection between these form-factors and the mass-dependent beta functions
of several gravitational couplings which include the Newton's constant as the most recent result.
All the beta functions that we have presented depend on a scale $q^2$
that is associated to the momentum dependence of the form-factors in Fourier space.
The running displays two important limits:
in the ultraviolet the beta functions coincide with their $\overline{\rm MS}$ counterparts,
while in the infrared the same beta functions go to zero with the leading power $\frac{q^2}{m^2}$
as expected from the Appelquist-Carazzone theorem.
We expect that our derivation of the semiclassical effective action could have
some relevant repercussion in the context of cosmology or astrophysics,
as it predicts effective values for the Newton's constant in units of the particles' masses
which depend on a physical scale of the renormalization group.

Besides the effects of decoupling, several other interesting results have been presented in this review.
In fact we have discussed the pragmatic connection
that is made in two dimensions with the expectations of Zamolodchikov's theorem.
Furthermore, in four dimensions we have established some interesting link
between the renormalization of the $R$ and $\Box R$ operators,
which might have implications for some approaches to quantum gravity.
In particular, we have made some speculation regarding the utility of our framework
for the asymptotic safety conjecture of quantum gravity,
in which a consistent non-perturbative renormalization of four-dimensional Einstein-Hilbert gravity
is assumed.

\bigskip

\noindent\emph{Acknowledgements.}
The research of O.Z.\ was funded by Deutsche Forschungsgemeinschaft (DFG) under the Grant Za~958/2-1.
T.P.N.\ acknowledges support from CAPES through the PNPD program.
S.A.F.\ acknowledges support from  the DAAD and the Ministerio de Educaci\'on Argentino under the ALE-ARG program.
O.Z.\ is grateful to Martin Reuter and all other partecipants of the workshop ``Quantum Fields -- From Fundamental Concepts to Phenomenological Questions''
for the interest shown in the topics of this work. The authors are grateful to Tiago~G.~Ribeiro and Ilya~L.~Shapiro
for collaborations on the projects discussed in this review, and to Carlo~Pagani for useful comments on the draft.

\appendix

\section{The non-local expansion of the heat kernel} \label{sect:heat-kernel}

The heat kernel of the Laplace-type operator ${\cal O}=\Delta_g + E$
is a bi-tensor that is defined as the solution of the differential equation
\begin{equation}
 \begin{split}
 \left(\partial_s + {\cal O}_x \right)  {\cal H}_D(s;x,x') =  0\,,  \qquad {\cal H}_D(0;x,x')=\delta^{(D)}(x,x')\,,
 \end{split}
\end{equation}
in which $\delta^{(D)}(x,x')$ is the covariant Dirac delta. The formal solution is the exponential
\begin{equation}
 \begin{split}
 {\cal H}_D(s;x,x') = \langle x | {\rm e}^{-s {\cal O}} | x '\rangle \,.
 \end{split}
\end{equation}
We keep the subscript $D$ as a reminder of the spacetime dimension for reasons explained in section \ref{sect:effective-action}.
A customary tool of quantum field theory is to consider the expression
\begin{equation}
 \begin{split}
 \ln (\frac{x}{y}) &= -\int_0^\infty  \frac{{\rm d}s}{s}\, \left( {\rm e}^{-s x}-{\rm e}^{-s y}\right)\,,
 \end{split}
\end{equation}
and use it to give a practical representation of the one loop functional trace
\begin{equation}
 \begin{split}
 \Ga[g] &= -\frac{1}{2} \tr \int_0^\infty \frac{{\rm d}s}{s} \int{\rm d}^Dx ~ {\rm e}^{-sm^2} {\cal H}_D(s;x,x)
 \end{split}
\end{equation}
modulo a field-independent normalization, as shown in the main text in \eqref{eq:effective-action-divergent}.

The heat kernel of a Laplace-type operator admits an expansion in powers of $s$ that starts with the power $s^{-D/2}$ known as Seeley-deWitt expansion.
The Seeley-deWitt expansion is perfectly suited for the computation of the divergences of the effective action,
and therefore for their $\overline{\rm MS}$ renormalization, but much less effective in obtaining the finite contributions of the effective action
that we need in this work.
As an alternative we consider
the non-local expansion of the heat kernel \cite{bavi87,bavi90,Codello:2012kq}. This latter expansion is a special curvature expansion known to the third order
that is valid for asymptotically free spacetimes and in which the effects of covariant derivatives are resummed.
The trace of the coincidence limit to the second order in the curvatures is
\begin{multline}
 {\cal H}(s) = \frac{1}{(4\pi s)^{d/2}} \int {\rm d}^D x \sqrt{g}\, {\rm tr} \Bigl\{
  \mathbf{1}
  + s G_E(s\Delta_g) E
  + s G_R(s\Delta_g) 
  +s^2 R F_R(s\Delta_g)R \\
  +s^2 R^{\mu\nu} F_{Ric}(s\Delta_g)R_{\mu\nu}
  +s^2 E F_E(s\Delta_g)E
  +s^2 E F_{RE}(s\Delta_g)R
  +s^2 \Omega^{\mu\nu} F_\Omega(s\Delta_g) \Omega_{\mu\nu}
 \Bigr\}
 + {\cal O}\left({\cal R}\right)^3\,,
\end{multline}
in which ${\cal O}\left({\cal R}\right)^3$ represents all possible non-local terms with three or more curvatures as described in \cite{bavi87,bavi90}.
The functions of $\Delta_g$ are known as form-factors of the heat kernel: they act on the rightmost curvature and should be regarded as non-local functions of the Laplacian.
The form-factors appearing in the linear terms have been derived in \cite{Codello:2012kq} as
\begin{equation}
 \begin{split}
 G_E(x) = -f(x)\,,\qquad G_R(x) = \frac{f(x)}{4}+\frac{f(x)-1}{2x}\,,
 \end{split}
\end{equation}
while those appearing in the quadratic terms have been derived in \cite{bavi87,bavi90} as
\begin{equation}
 \begin{split}
 &F_{Ric}(x) = \frac{1}{6x}+\frac{f(x)-1}{x^2} \,\qquad
 F_R(x) = -\frac{7}{48x}+\frac{f(x)}{32}+\frac{f(x)}{8x}-\frac{f(x)-1}{8x^2} \\
 &F_{RE}(x) = -\frac{f(x)}{4}-\frac{f(x)-1}{2x} \,\qquad
 F_E(x) =\frac{f(x)}{2} \,\qquad
 F_\Omega(x) = -\frac{f(x)-1}{2x}\,,
 \end{split}
\end{equation}
but we give them in the notation of \cite{Codello:2012kq}.
Interestingly all the above form-factors depend on a basic form-factor which is defined as
\begin{equation}
 \begin{split}
 f(x) &= \int_0^1 \!{d}\alpha \, {\rm e}^{-\alpha(1-\alpha)x}\,.
 \end{split}
\end{equation}
All the form-factors admit well-defined expansions both for large and for small values of the parameter $s$ \cite{bavi87,bavi90}
and therefore allow us to go beyond the simple asymptotic expressions at small $s$.

\section{Further mathematical details} \label{sect:further}

We collect here some useful formulas for dealing with simplifications of the curvature tensors and the Dirac operator that are used in sections \ref{sect:nonlocal-two} and \ref{sect:nonlocal-four}.
In $D=2$ all Riemaniann curvature tensors can be written in terms of the metric and the curvature scalar $R$ because only the conformal factor of the metric is an independent
degree of freedom. The Riemann and the Ricci tensors are simplified as
\beq
R_{\mu\nu\al\be} =  \frac12\,R\big(g_{\mu\al}g_{\nu\be} -
g_{\nu\al}g_{\mu\be}\big),
\qquad
R_{\mu\nu} =  \frac12\,R\,g_{\mu\nu}\,.
\eeq
Notice that in \eqref{eq:effective-action-2d} we use explicitly the above formulas to argue that the only
relevant quadratic form-factor in $D=2$ involves two copies of the scalar curvature. As discussed in section \ref{sect:effective-action}
we have continued the dimensionality only through the dependence of the leading power of the heat kernel and all geometric tensors behave
as if they live in precisely two dimensions, which allows us to use the above simplifications.
In $D=4$ instead all curvature tensors are generally independent and for \eqref{eq:effective-action-full}
we have chosen a basis that includes the Ricci scalar and the Weyl tensor,
which is useful to disentangle the contributions coming from the conformal factor from those of purely spin-$2$ parts of $g_{\mu\nu}$ that are missing in $D=2$.

Our conventions for the Dirac operator are in form the same for both $D=2$ and $D=4$.
The spin connection $\omega_\mu {}^{a}{}_{b}$ is constructed from the Levi-Civita connection in a straightforward way
by introducing the $D$-bein $e^a{}_\mu$ that trivialize the metric $g_{\mu\nu} =e^a{}_\mu e^b{}_\nu \delta_{ab}$,
and requiring the compatibility of the extended connection $\nabla_\mu e^a{}_\nu=0$.
We use the fact that the elements $\sigma_{ab}=\frac{i}{2}\left[\gamma_a,\gamma_b\right]$ of the Clifford algebra are generators of local Lorentz transformations
to construct the covariant connection acting on Dirac fields
$$D_\mu = \partial_\mu - \frac{i}{4}\omega_\mu {}^{ab}\sigma_{ab}\,,$$
which appears in \eqref{eq:bare-actions}.
When applying the general formulas for the heat kernel we need the curvature two-form on Dirac fields
\begin{eqnarray}
\Omega_{\mu\nu} = \left[D_\mu,D_\nu\right]
 = - \frac{i}{4}F_{\mu\nu} {}^{ab}\sigma_{ab}
\end{eqnarray}
in which $F_{\mu\nu} {}^{ab} = R_{\mu\nu}{}^{\rho\theta}e^a{}_\rho e^b{}_\theta$ is the spin curvature of $\omega_\mu {}^{a}{}_{b}$.
Using some standard properties of the Clifford algebra, we explicitly find
\begin{eqnarray}
\tr ~\Omega^2 = -\frac{d_\gamma}{8} R_{\mu\nu\rho\theta} R^{\mu\nu\rho\theta}
\end{eqnarray}
in which $d_\gamma = \tr \, {\mathbf 1}$ is the dimensionality of the Clifford algebra. Interestingy $d_\gamma$
factorizes from all formulas of sections \ref{sect:dirac2d} and \ref{sect:dirac4d}
because our bare actions are invariant under chiral symmetry
signalling the fact that it is the product $n_{\rm f}\cdot d_\gamma$ that effectively counts the number of independent fermionic degrees of freedom.

\end{document}